\documentclass{article}
\usepackage{amsfonts}

\input{tcilatex}

\begin{document}

\title{Electromagnetism on Anisotropic Fractals}
\author{\textbf{Martin Ostoja-Starzewski} \\
%EndAName
Department of Mechanical Science and Engineering\\
University of Illinois at Urbana-Champaign\\
Urbana, IL 61801, U.S.A.}
\maketitle

\begin{abstract}
We derive basic equations of electromagnetic fields in fractal media which
are specified by three indepedent fractal dimensions $\alpha _{i}$ in the
respective directions $x_{i}$ ($i=1,2,3$) of the Cartesian space in which
the fractal is embedded.\ To grasp the generally anisotropic structure of a
fractal, we employ the product measure, so that the global forms of
governing equations may be cast in forms involving conventional
(integer-order) integrals, while the local forms are expressed through
partial differential equations with derivatives of integer order but
containing coefficients involving the $\alpha _{i}$'s.\ First, a formulation
based on product measures is shown to satisfy the four basic identities of
vector calculus. This allows a generalization of the Green-Gauss and Stokes
theorems as well as the charge conservation equation on anisotropic
fractals. Then, pursuing the conceptual approach, we derive the Faraday and
Amp\`{e}re laws for such fractal media, which, along with two auxiliary
null-divergence conditions, effectively give the modified Maxwell equations.
Proceeding on a separate track, we employ a variational principle for
electromagnetic fields, appropriately adapted to fractal media, to
independently derive the same forms of these two laws. It is next found that
the parabolic (for a conducting medium) and the hyperbolic (for a dielectric
medium) equations involve modified gradient operators, while the Poynting
vector has the same form as in the non-fractal case. Finally, Maxwell's
electromagnetic stress tensor is reformulated for fractal systems. In all
the cases, the derived equations for fractal media depend explicitly on
fractal dimensions and reduce to conventional forms for continuous media
with Euclidean geometries upon setting the dimensions to integers.
\end{abstract}

\setcounter{equation}{0}

\section{Introduction}

In a recent article (Li \& Ostoja-Starzewski, 2009) we formulated a
continuum mechanics of anisotropic fractal elastic media using two
independent approaches: a mechanical (Newtonian) one and a variational
(Lagrangian-Hamiltonian) one. Proceeding separately on both paths, we
derived the identical equations of wave motion in one-, two- and
three-dimensional media, and this provided a verification of the consistency
of the product measure capable of grasping the anisotropic fractal media as
well as the verification of the final equations. The question arises whether
one can use the same strategy to derive the field equations of
electromagnetism in fractal media, be they isotropic or anisotropic. Various
studies generalizing the classical, quantum or relativistic field theories
to fractal spaces (Stillinger, 1977; Svozil, 1987; Palmer \&\ Stavrinou,
2004; Nottale, 2010) as well as the electromagnetism to fractal media
appeared over the past two decades (Tarasov, 2006, 2010). They have been
formulated either through a conceptual path $-$ in the case of
electromagnetism, by developing the Gauss, Faraday and Amp\`{e}re laws, so
as to obtain the Maxwell equations $-$ or through variational principles.
However, a derivation obtained on one path (say, conceptual) has never been
verified through another, separate derivation (respectively, variational).

In the following we proceed under the premise that the conceptual derivation
of Maxwell's equations for anisotropic fractal media should yield the same
results as the derivation based on a variational principle, both derivations
being based on the same dimensional regularization. To grasp the generally
anisotropic structure of a fractal, we employ the recently introduced
product measure, so that the global forms of governing equations may be cast
in forms involving conventional (integer-order) integrals, while the local
forms are expressed through partial differential equations with derivatives
of integer order but containing coefficients involving fractal dimensions $%
\alpha _{i}$ in three respective directions $x_{i},$ $i=1,2,3$.

We first show that the formulation based on product measures satisfies the
four basic identities of vector calculus. This readily leads to a
generalization of the Green-Gauss and Stokes theorems involving the fractal
gradient and curl operators, and hence to the charge conservation equation
on anisotropic fractals. Then, pursuing the conceptual approach, we derive
the Faraday and Amp\`{e}re laws for such fractal media, which, along with
two auxiliary null-divergence conditions, effectively give the modified
Maxwell equations.

Proceeding on a separate track, we employ a variational principle for
electromagnetic fields, appropriately adapted to fractal media, to
independently derive the same forms of these two laws. We next examine
whether the parabolic (for a conducting medium) and the hyperbolic (for a
dielectric medium) equations involve fractal gradient operators, and also
whether the Poynting vector has the same form as in the non-fractal case.
Finally, Maxwell's electromagnetic stress tensor is reformulated for fractal
systems. In all the cases, the derived equations for fractal media depend
explicitly on fractal dimensions and reduce to conventional forms for
continuous media with Euclidean geometries upon setting the dimensions to
integers. In order to make the presentation clear, in some places we give
the tensorial relations in the symbolic as well as the index notations.

\setcounter{equation}{0}

\section{Background}

\subsection{Anisotropic Fractal Distributions}

The distribution of charges in an electrically and magnetically conducting
isotropic fractal medium embedded in the Euclidea space $\mathbb{E}^{3}$
obeys a power law%
\begin{equation}
Q(R)\sim R^{D},\text{ \ \ \ }D<3,
\end{equation}%
where $D$ is the fractal dimension and $R$ is the spatial resolution.
However, in an anisotropic fractal medium (generally denoted by $\mathcal{W}$%
) we write%
\begin{equation}
Q(x_{1},x_{2},x_{3})\sim x_{1}{}^{\alpha _{1}}x_{2}{}^{\alpha
_{2}}x_{3}{}^{\alpha _{3}},
\end{equation}%
where each $\alpha _{k}$ plays the role of a fractal dimension along each
axis $x_{k}$ ($i=1,2,3$) of the Cartesian space in which the fractal is
embedded. In other directions, the fractal dimension is not necessarily the
sum of projected fractal dimensions, however, as noted by\ Falconer (2003), "%
\textit{Many fractals encountered in practice are not actually products, but
are product-like}." Hence, we expect the equality between the fractal
dimension $D$ of the total charge and $\alpha _{1}+\alpha _{2}+\alpha _{3}$
to hold for fractals encountered in practice. When $D\rightarrow 3$, with
each $\alpha _{i}\rightarrow 1$, the conventional concept of charge
distribution is recovered.

By analogy to the formulation of continuum mechanics on anisotropic fractals
(Li \&\ Ostoja-Starzewski, 2009), we now express (2.1) through a product
measure%
\begin{equation}
Q(\mathcal{W})=\int_{\mathcal{W}}\rho (x_{1},x_{2},x_{3})dl_{\alpha
_{1}}(x_{1})dl_{\alpha _{2}}(x_{2})dl_{\alpha _{3}}(x_{3}),
\end{equation}%
where $\rho $\ is the density of distribution. In (2.3), the length measure
in each coordinate is given in terms of the transformation coefficients $%
c_{1}^{(k)}$ by 
\begin{equation}
dl_{\alpha _{k}}(x_{k})=c_{1}^{(k)}(\alpha _{k},x_{k})dx_{k},\text{ \ \ }%
k=1,2,3\text{ \ \ \ (no sum),}
\end{equation}%
where $l_{k}$ is the total length (integral interval) along $x_{k}$ and $%
l_{k0}$\ is the characteristic length in the given direction, like the mean
pore size. The relation (2.4) implies that the infinitesimal fractal volume
element, $dV_{D}$, is related to the cubic volume element, $%
dV_{3}=dx_{1}dx_{2}dx_{3}$ by \begingroup\renewcommand{\arraystretch}{1.9} 
\begin{equation}
\begin{array}{c}
\displaystyle dV_{D}=dl_{\alpha _{1}}(x_{1})dl_{\alpha
_{2}}(x_{2})dl_{\alpha
_{3}}(x_{3})=c_{1}^{(1)}c_{1}^{(2)}c_{1}^{(3)}dx_{1}dx_{2}dx_{3}=c_{3}dV_{3},
\\ 
\displaystyle\text{with \ \ }c_{3}=c_{1}^{(1)}c_{1}^{(2)}c_{1}^{(3)}\text{.}%
\end{array}%
\end{equation}%
\endgroup Similarly, the infinitesimal fractal surface element, $dS_{d}$, is
related to the planar surface element, $dS_{d}^{(k)}=dx_{i}dx_{j}$ with the
normal vector along $x_{k}$, according to%
\begin{equation}
\begin{array}{c}
\displaystyle dS_{d}^{(k)}=c_{c}dS_{2}, \\ 
\displaystyle\text{with \ \ }c_{2}^{\left( k\right)
}=c_{1}^{(i)}c_{1}^{(j)}=c_{3}/c_{1}^{(k)},\text{ \ \ \ }i\neq j,\text{ \ \
\ }i,j\neq k.%
\end{array}%
\end{equation}%
The sum $d^{\left( k\right) }=\alpha _{i}+\alpha _{j},$\ $i\neq j,$ $i,j\neq
k$, is the fractal dimension of the surface $S_{d}^{(k)}$ along the
diagonals $|x_{i}|=|x_{j}|$ in $S_{d}^{(k)}$. Again, this equality is not
necessarily true elsewhere, but is expected to hold for fractals encountered
in practice (Falconer, 2003).

The transformation coefficients $c_{1}^{(k)}$ showing up in (2.4) can be
represented in terms of the modified Riemann-Liouville fractional integral
of Jumarie (2005, 2009), thus%
\begin{equation}
c_{1}^{(k)}=\alpha _{k}\left( \frac{l_{k}-x_{k}}{l_{k0}}\right) ^{\alpha
_{k}-1},\text{ \ \ }k=1,2,3,\text{ \ \ \ (no sum),}
\end{equation}%
although further developments will not depend on this explicit form.

\subsection{Vector calculus on anisotropic fractals}

Following (Li \& Ostoja-Starzewski, 2009), we will extensively employ the 
\textit{fractal derivative} (\textit{fractal gradient}) operator ($\func{grad%
}$)%
\begin{equation}
\mathbf{\nabla }^{D}\phi =\mathbf{e}_{k}\nabla _{k}^{D}\phi \text{ \ \ \ or
\ \ \ }\nabla _{k}^{D}\phi =\frac{1}{c_{1}^{(k)}}\frac{\partial \phi }{%
\partial x_{k}}\text{ \ (no sum on }k\text{)}.
\end{equation}%
where $\mathbf{e}_{k}$\ are base vectors. Hence, the \textit{fractal
divergence} of a vector field%
\begin{equation}
\func{div}\mathbf{f}=\mathbf{\nabla }^{D}\cdot \mathbf{f}\text{ \ \ \ or \ \
\ }\nabla _{k}^{D}f_{k}=\frac{1}{c_{1}^{(k)}}\frac{\partial f_{k}}{\partial
x_{k}}.
\end{equation}%
Note that this leads to a \textit{fractal curl} operator of a vector field%
\begin{equation}
\func{curl}\mathbf{f}=\mathbf{\nabla }^{D}\times \mathbf{f}\text{ \ \ \ or \
\ \ }\nabla _{k}^{D}\times f_{i}=e_{jki}\frac{1}{c_{1}^{(k)}}\frac{\partial
f_{i}}{\partial x_{k}}.
\end{equation}

We observe that the four fundamental identities of the conventional vector
calculus carry over to these new operators:

(i) The divergence of the curl of a vector field $\mathbf{f}$: 
\begin{equation}
\func{div}\cdot \func{curl}f=\nabla _{k}^{D}\times \mathbf{f}=\frac{1}{%
c_{1}^{(j)}}\frac{\partial f}{\partial x_{j}}\left[ e_{jki}\frac{1}{%
c_{1}^{(k)}}\frac{\partial f_{i}}{\partial x_{k}}\right] =e_{jki}\frac{1}{%
c_{1}^{(j)}}\frac{1}{c_{1}^{(k)}}\frac{\partial f_{i}}{\partial
x_{j}\partial x_{k}}=0.
\end{equation}

(ii) The curl of the gradient of a scalar field $\phi $: 
\begin{equation}
\func{curl}\times (\func{grad}\phi )=\nabla _{j}^{D}\times (\nabla
_{k}^{D}\times \phi )=e_{ijk}\frac{1}{c_{1}^{(j)}}\frac{\partial }{\partial
x_{j}}\left[ \frac{1}{c_{1}^{(k)}}\frac{\partial \phi }{\partial x_{k}}%
\right] =e_{jki}\frac{1}{c_{1}^{(j)}}\frac{1}{c_{1}^{(k)}}\frac{\partial
f_{i}}{\partial x_{j}\partial x_{k}}=0.
\end{equation}%
In both cases above we can pull $1/c_{1}^{(k)}$\ in front of the gradient
because the coefficient $c_{1}^{(k)}$ is independent of $x_{j}$.

(iii) The divergence of the gradient of a scalar field $\phi $ is written in
terms of the fractal gradient as%
\begin{equation}
\func{div}\cdot (\func{grad}\phi )=\nabla _{j}^{D}\cdot \nabla _{k}^{D}\phi =%
\frac{1}{c_{1}^{(j)}}\frac{\partial }{\partial x_{j}}\left[ \frac{1}{%
c_{1}^{(j)}}\frac{\partial \phi }{\partial x_{j}}\right] =\frac{1}{%
c_{1}^{(j)}}\left[ \frac{\partial \phi ,_{j}}{c_{1}^{(j)}}\right] ,_{j}
\end{equation}%
which gives an explicit form of the fractal Laplacian.

(iv) The curl of the curl operating on a vector field $\mathbf{f}$:%
\begingroup\renewcommand{\arraystretch}{2.2}%
\begin{equation}
\begin{array}{c}
\displaystyle\func{curl}\times (\func{curl}\mathbf{f})=\nabla _{r}^{D}\times
(\nabla _{r}^{D}\times \mathbf{f})=e_{prj}\frac{1}{c_{1}^{(r)}}\frac{%
\partial }{\partial x_{r}}\left[ e_{jki}\frac{1}{c_{1}^{(k)}}\frac{\partial
f_{i}}{\partial x_{k}}\right] \\ 
\displaystyle=e_{prj}e_{jki}\frac{1}{c_{1}^{(r)}}\frac{\partial }{\partial
x_{r}}\left[ \frac{1}{c_{1}^{(k)}}\frac{\partial f_{i}}{\partial x_{k}}%
\right] =(\delta _{kp}\delta _{ir}-\delta _{kr}\delta _{ip})\frac{1}{%
c_{1}^{(r)}}\frac{\partial }{\partial x_{r}}\left[ \frac{1}{c_{1}^{(k)}}%
\frac{\partial f_{i}}{\partial x_{k}}\right] \\ 
\displaystyle=\frac{1}{c_{1}^{(r)}}\frac{\partial }{\partial x_{r}}\left[ 
\frac{1}{c_{1}^{(p)}}\frac{\partial f_{r}}{\partial x_{p}}\right] -\frac{1}{%
c_{1}^{(r)}}\frac{\partial }{\partial x_{r}}\left[ \frac{1}{c_{1}^{(r)}}%
\frac{\partial f_{p}}{\partial x_{r}}\right] =\nabla _{p}^{D}\left( \nabla
_{r}^{D}\cdot \mathbf{f}\right) -\nabla _{p}^{D}\nabla _{r}^{D}\mathbf{f}%
\end{array}%
\end{equation}%
\endgroup

As further background, we now extend two integral theorems to generally
anisotropic fractals.

\subsection{Stokes and Green-Gauss theorems for anisotropic fractals}

\textbf{Stokes theorem.} We begin with $\int_{A}(\mathbf{\nabla }^{D}\times 
\mathbf{f})\cdot \mathbf{n}dS_{d}$ and proceed in the index notation:%
\begingroup\renewcommand{\arraystretch}{2.2}%
\begin{equation}
\begin{array}{c}
\displaystyle\int_{\partial \mathcal{W}}n_{k}\nabla
_{j}^{D}f_{i}dS_{d}\equiv \int_{\partial \mathcal{W}}n_{k}e_{kji}\frac{1}{%
c_{1}^{(j)}}f_{i},_{j}dS_{d}=\int_{\partial \mathcal{W}}n_{k}e_{kji}\frac{1}{%
c_{1}^{(j)}}f_{i},_{j}c_{2}^{(k)}dS_{2} \\ 
\displaystyle=\int_{\partial \mathcal{W}%
}n_{k}e_{kji}f_{i},_{j}c_{1}^{(i)}dS_{2}=\int_{\partial \mathcal{W}%
}n_{k}e_{kji}\left( f_{i}c_{1}^{(i)}\right) ,_{j}dS_{2}%
\end{array}%
\end{equation}%
\endgroup In the second equality above we employed the dimensional
regularization, followed by taking note of (2.6), including the fact that $%
c_{1}^{(i)}$\ is independent of $x_{j}$. Next, by the Stokes theorem in the
Euclidean space $\mathbb{E}^{3}$, we have%
\begin{equation}
\int_{\partial \mathcal{W}}n_{k}e_{kji}\left( f_{i}c_{1}^{(i)}\right)
,_{j}dS_{2}=\int_{\partial \mathcal{W}}f_{i}c_{1}^{(i)}dl_{i}=%
\int_{A}f_{i}dl_{iD}
\end{equation}%
where, again, we used (2.4). Thus, we arrive at the \textit{Stokes theorem
for fractals}%
\begin{equation}
\int_{A}\mathbf{n}\cdot \func{curl}\mathbf{f}\text{ }dS_{d}=\int_{l}\mathbf{f%
}\cdot d\mathbf{l}_{\alpha _{1}}
\end{equation}%
Formally, the above agrees with Tarasov's (2006) result in the case of
isotropy, although our $\func{curl}$ operator\ also grasps anisotropy.

\textbf{Green-Gauss theorem.} We begin with $\int_{A}\mathbf{f\otimes n}%
dS_{d}$ and proceed in the index notation:\begingroup\renewcommand{%
\arraystretch}{2.2}%
\begin{equation}
\begin{array}{c}
\displaystyle\int_{\partial \mathcal{W}}f_{i}n_{k}dS_{d}=\int_{\partial 
\mathcal{W}}f_{i}n_{k}c_{2}^{(k)}dS_{2}=\int_{\mathcal{W}}\left(
f_{i}c_{2}^{(k)}\right) ,_{k}dV_{2} \\ 
\displaystyle=\int_{\mathcal{W}}\left( f_{i}c_{2}^{(k)}\right)
,_{k}c_{3}^{-1}dV_{D}=\int_{\mathcal{W}%
}f_{i},_{k}c_{2}^{(k)}c_{3}^{-1}dV_{D}=\int_{\mathcal{W}}f_{i},_{k}\frac{1}{%
c_{1}^{(k)}}dV_{D},%
\end{array}%
\end{equation}%
\endgroup whereby in the second equality we employed the dimensional
regularization, followed by taking note of (2.5), then followed by employing
the Green-Gauss theorem in $\mathbb{E}^{3}$, and in turn followed by noting
that $c_{2}^{(k)}$\ is independent of $x_{k}$. Thus, we arrive at the 
\textit{Green-Gauss theorem for fractals} as%
\begin{equation}
\int_{\partial \mathcal{W}}\mathbf{f\cdot n}\text{ }dS_{d}=\int_{\mathcal{W}%
}\nabla ^{D}\mathbf{f}\text{ }dV_{D}
\end{equation}%
The same comment as that following (2.7) applies here.

\subsection{Charge conservation on anisotropic fractals}

Let $\mathbf{J}$ be the current density, $\mathbf{n}$\ be the direction
normal to the surface, and $\eta $\ be the charge density on the fractal $%
\mathcal{W}$. Then we have%
\begin{equation}
\int_{\partial \mathcal{W}}\mathbf{J\cdot n}\text{ }dS_{d}=-\int_{\mathcal{W}%
}\eta \text{ }dV_{D}
\end{equation}%
On account of (2.18), we obtain%
\begin{equation}
\begin{array}{cc}
\text{global form} & \displaystyle\int_{\mathcal{W}}\nabla ^{D}\cdot \mathbf{%
J}\text{ }dV_{D}=-\frac{d}{dt}\int_{\mathcal{W}}\eta \text{ }dV_{D} \\ 
\text{local form} & \displaystyle\nabla ^{D}\cdot \mathbf{J}\text{ }=-\frac{%
\partial }{\partial t}\eta \text{ }%
\end{array}%
\end{equation}%
The fact that $\nabla ^{D}$\ automatically appears in the above will be
exploited below.

Focusing on a linear electromagnetic response, with $\mathbf{\sigma }$\
being the electric conductivity tensor, Ohm's law for an anisotropic medium
reads%
\begin{equation}
\mathbf{J}=\mathbf{\sigma }\cdot \mathbf{E}\text{ \ \ \ or \ \ \ }%
J_{i}=\sigma _{ij}E_{j}
\end{equation}%
the isotropic form following for $\sigma _{ij}=\delta _{ij}\sigma $. The
fact that the constitutive form above carries over directly from the
non-fractal case is motivated by the analogy to elastic media, where Hooke's
law is unchanged when going from non-fractal to fractal media (Li \&\
Ostoja-Starzewski, 2009); that result ensured the consistency of the
Newtonian and Hamiltonian approaches to the derivation of governing
equations - recall the beginning of the Introduction.

\setcounter{equation}{0}

\section{Formulation of Maxwell's equations via Stokes and Green-Gauss
theorems}

\subsection{Faraday's law}

First, note that the Faraday law holds for a fractal $\mathcal{W}$ embedded
in $\mathbb{E}^{3}$

\begin{equation}
\frac{d}{dt}\int_{A}\mathbf{B}\cdot \mathbf{n}dA_{d}=-\int_{l}\mathbf{E}%
\cdot d\mathbf{l}_{D}
\end{equation}%
On account of the Stokes theorem for fractals (2.17), we obtain%
\begin{equation}
\frac{d}{dt}\int_{A}B_{k}n_{k}dA_{d}=-\frac{d}{dt}\int_{A}e_{kji}\nabla
_{j}^{D}E_{i}n_{k}dA_{d}\equiv -\frac{d}{dt}\int_{A}e_{kji}\frac{1}{%
c_{1}^{(j)}}E_{i},_{j}n_{k}dA_{d}
\end{equation}%
which, by localization leads, to\begingroup\renewcommand{\arraystretch}{2.2}%
\begin{equation}
\begin{array}{c}
\displaystyle0=\frac{\partial }{\partial t}B_{k}+e_{kji}\nabla
_{j}^{D}E_{i}\equiv \frac{\partial }{\partial t}B_{k}+e_{kji}\frac{1}{%
c_{1}^{(j)}}E_{i},_{j} \\ 
\text{or} \\ 
\displaystyle\mathbf{0}=\frac{\partial }{\partial t}\mathbf{B}+\mathbf{%
\nabla }^{D}\times \mathbf{E}%
\end{array}%
\end{equation}%
\endgroup

\subsection{Amp\`{e}re's law}

First, note that the Amp\`{e}re law holds for a fractal $\mathcal{W}$
embedded in $\mathbb{E}^{3}$

\begin{equation}
\int_{A}\mathbf{C}\cdot \mathbf{n}dA_{d}=\int_{l}\mathbf{H}\cdot d\mathbf{l}%
_{D}=\int_{l}\mathbf{n}\cdot (\mathbf{\nabla }^{D}\times \mathbf{H})dA_{d}
\end{equation}%
Noting%
\begin{equation}
\mathbf{C}=\mathbf{J}+\frac{\partial \mathbf{D}}{\partial t}
\end{equation}%
we obtain%
\begin{equation}
\int_{A}\left( \mathbf{J}+\frac{\partial \mathbf{D}}{\partial t}-\mathbf{%
\nabla }^{D}\times \mathbf{H}\right) dA_{d}=0
\end{equation}%
which, again by focusing on the integrand, yields%
\begin{equation}
\mathbf{J}+\frac{\partial \mathbf{D}}{\partial t}-\mathbf{\nabla }^{D}\times 
\mathbf{H}=\mathbf{0}
\end{equation}%
On account of the speed of light being related to the dielectric ($\epsilon
_{0}$) and magnetic ($\mu _{0}$) constants by the well known relation $c=1/%
\sqrt{\epsilon _{0}\mu _{0}}$, the above is equivalent to%
\begin{equation}
\frac{1}{\epsilon _{0}}\mathbf{J}+\frac{\partial \mathbf{E}}{\partial t}%
-c^{2}\mathbf{\nabla }^{D}\times \mathbf{B}=\mathbf{0}
\end{equation}

The equations (3.7) and (3.8) are subject to the constraints%
\begin{equation}
\begin{array}{cc}
\text{Gauss law for magnetism} & \nabla ^{D}\cdot \mathbf{B}=0 \\ 
\text{Gauss law} & \nabla ^{D}\cdot \mathbf{E}=0%
\end{array}%
\end{equation}%
where the presence of the fractal (rather than the classical) divergence
operator $\nabla ^{D}$ has clearly been suggested by (2.21).

\setcounter{equation}{0}

\section{Derivation from variational principle}

The variational principle for the Maxwell equations in conventional
(non-fractal) setting $\mathbb{E}^{3}$\begingroup\renewcommand{%
\arraystretch}{2.2}%
\begin{equation}
\begin{array}{cc}
\nabla \cdot \mathbf{E}=0, & \nabla \cdot \mathbf{B}=0, \\ 
\displaystyle\frac{\partial \mathbf{B}}{\partial t}+\nabla \times \mathbf{E}%
=0, & \displaystyle\frac{\partial \mathbf{E}}{\partial t}-c^{2}\nabla \times 
\mathbf{B}+\frac{1}{\epsilon _{0}}\mathbf{J}=0,%
\end{array}%
\end{equation}%
\endgroup is (Seliger \&\ Whitham, 1968)\begingroup\renewcommand{%
\arraystretch}{2.2}%
\begin{equation}
\begin{array}{c}
\displaystyle\delta \int \int \mathcal{L}dV_{3}dt=0, \\ 
\text{with} \\ 
\displaystyle\mathcal{L}=\epsilon _{0}\left[ \frac{1}{2}c^{2}B_{i}B_{i}-%
\frac{1}{2}E_{i}E_{i}+A_{i}\left( \frac{\partial E_{i}}{\partial t}%
-c^{2}e_{ijk}B_{k},_{j}+\frac{1}{\epsilon _{0}}J_{i}\right) +\chi E_{i},_{i}%
\right] .%
\end{array}%
\end{equation}%
\endgroup Here the vector potential $\mathbf{A}$\ and the scalar potential $%
\chi $\ are the Lagrange multipliers, while the integrand of (4.2) is
slightly extended to include the current density $\mathbf{J}$. By varying
(4.2)$_{1}$ with respect to $\mathbf{E}$, $\mathbf{B}$, $\mathbf{A}$, and $%
\chi $, gives, respectively,\begingroup\renewcommand{\arraystretch}{1.8}%
\begin{equation}
\begin{array}{cc}
\delta \mathbf{E}: & \displaystyle\mathbf{E=-\partial A}/\partial t-\mathbf{%
\nabla }\chi \\ 
\delta \mathbf{B}: & \mathbf{B=\nabla \times A} \\ 
\delta \mathbf{A}: & \displaystyle\mathbf{\partial E}/\partial t-c^{2}%
\mathbf{\nabla }\times \mathbf{B}+\frac{1}{\epsilon _{0}}\mathbf{J}=0 \\ 
\delta \chi : & \mathbf{\nabla }\cdot \mathbf{E}=0%
\end{array}%
\end{equation}%
\endgroup It is clear that (4.3)$_{1,2}$ satisfy the first pair of Maxwell's
equations (4.1) identically. Henceforth, our goal is to obtain the Maxwell
equations (4.1)$_{1}$-(4.1)$_{4}$ for an anisotropic fractal, by using the
same type of a variational approach.

For a fractal $\mathcal{W}$ embedded in $\mathbb{E}^{3}$, the variational
principle for electromagnetic fields is written as\begingroup%
\renewcommand{\arraystretch}{2.2}%
\begin{equation}
\delta \int_{t_{1}}^{t_{2}}\int_{\mathcal{W}}\mathcal{L}dV_{D}dt=0
\end{equation}%
with $\mathcal{L}$\ given by (4.2)$_{2}$.

Varying (4.4) with respect to $\mathbf{E}$, we obtain%
\begin{equation}
\mathbf{E}=\mathbf{-\partial A}/\partial t-\mathbf{\nabla }^{D}(\chi c_{1})%
\text{ \ \ \ \ or \ \ \ \ }E_{i}=-\mathbf{\partial }A_{i}/\partial t-\frac{1%
}{c_{1}^{(i)}}\left( \chi c_{1}^{(i)}\right) ,_{i}\equiv \nabla
_{i}^{D}\left( A_{i}c_{1}^{(i)}\right)
\end{equation}%
while varying it with respect to $\mathbf{B}$, we obtain the Gauss law for
magnetism on fractals%
\begin{equation}
\mathbf{B}=\mathbf{\nabla }^{D}\times \left( \mathbf{A}c_{1}\right) \text{ \
\ \ \ or \ \ \ \ }B_{k}=e_{ijk}\frac{1}{c_{1}^{(j)}}\left(
A_{i}c_{1}^{(j)}\right) ,_{j}\equiv \nabla _{j}^{D}\left(
A_{i}c_{1}^{(j)}\right)
\end{equation}%
In both cases, the fractal gradient operator, $\mathbf{\nabla }^{D}$, shows
up due to the presence of $dV_{D}$\ in (4.4)$_{1}$; recall (2.5)$_{1}$.

The next step is to verify whether substituting (4.5) into (3.3) would give
a zero vector. Thus,\begingroup\renewcommand{\arraystretch}{2.2}%
\begin{equation}
\begin{array}{c}
\displaystyle0=\frac{\partial }{\partial t}B_{k}+e_{kji}\frac{1}{c_{1}^{(j)}}%
E_{i},_{j}=\mathbf{\partial }B_{k}/\partial t+e_{kji}\frac{1}{c_{1}^{(i)}}%
\left[ -\frac{\partial }{\partial t}\left( A_{i}c_{1}^{(i)}\right) -\left(
\chi c_{1}^{(i)}\right) ,_{i}\right] ,_{j} \\ 
\displaystyle=\mathbf{\partial }B_{k}/\partial t+e_{kji}\frac{1}{c_{1}^{(i)}}%
\left[ -\frac{\partial }{\partial t}\left( A_{i}c_{1}^{(i)}\right)
,_{j}-\left( \chi c_{1}^{(i)}\right) ,_{ij}\right] \\ 
\displaystyle=\frac{\partial }{\partial t}\left[ B_{k}-e_{kji}\frac{1}{%
c_{1}^{(i)}}\left( A_{i}c_{1}^{(i)}\right) ,_{j}\right] -e_{kji}\frac{1}{%
c_{1}^{(i)}}\left( \chi c_{1}^{(i)}\right) ,_{ij} \\ 
\displaystyle\equiv \frac{\partial }{\partial t}\left[ B_{k}-\nabla
_{j}^{D}\times \left( A_{i}c_{1}^{(i)}\right) \right] -\nabla _{j}^{D}\times
\left( \chi c_{1}^{(i)}\right) ,_{i}%
\end{array}%
\end{equation}%
\endgroup The first term in the square brackets in the last line of the
above agrees with (4.6), while the last term vanishes, providing the fractal
is isotropic ($c_{1}^{(i)}=c_{1}^{(j)}$, $i\neq j$), in which case (4.7) is
satisfied identically. Thus, for an anisotropic fractal, there appears a
source/disturbance $-$ an observation consistent with Tarasov's formulation,
though his explicit form differs from ours.

Also, substituting (4.6) into (3.9)$_{1}$ yields%
\begin{equation}
0=e_{ijk}\frac{1}{c_{1}^{(k)}}\frac{1}{c_{1}^{(j)}}\left(
A_{i}c_{1}^{(j)}\right) ,_{jk}\equiv \nabla _{k}^{D}\nabla _{j}^{D}\times
\left( A_{i}c_{1}^{(j)}\right)
\end{equation}%
which vanishes providing the fractal is isotropic ($c_{1}^{(i)}=c_{1}^{(j)}$%
, $i\neq j$).

Varying (4.4) with respect to $\mathbf{A}$, we obtain\begingroup%
\renewcommand{\arraystretch}{2.2}%
\begin{equation}
\mathbf{\partial E}/\partial t-c^{2}\mathbf{\nabla }^{D}\times \mathbf{B}+%
\frac{1}{\epsilon _{0}}\mathbf{J}=\mathbf{0}\text{ \ \ \ \ or \ \ \ \ }%
\mathbf{\partial }E_{i}/\partial t-c^{2}\nabla _{j}^{D}\left(
B_{k}c_{1}^{(j)}\right) +\frac{1}{\epsilon _{0}}J_{i}=0
\end{equation}%
\endgroup which agrees with (3.8) perfectly.

Varying (4.4) with respect to $\chi $, we obtain the Gauss law%
\begin{equation}
\nabla ^{D}\cdot \mathbf{E}=0\text{ \ \ \ \ or \ \ \ \ }\nabla _{i}^{D}E_{i}=%
\frac{1}{c_{1}^{(i)}}E_{i},_{i}=0
\end{equation}%
which perfectly agrees with (3.9)$_{2}$. Thus, we have derived by an
entirely independent route the same set of Maxwell's equations modified to
fractals as those in Section 2. In analogy to (Li \&\ Ostoja-Starzewski,
2009) which treated the continuum mechanics of fractal elastic media
independently by Newtonian and Lagrangian approaches, this shows that the
formulation of continuum physics equations based on the product measures is
consistent. In the case of isotropy, our equations do not reduce to those of
Tarasov (2010).

\setcounter{equation}{0}

\section{Second order differential equations of electromagnetism}

In Gaussian units Eqns (3.9), (3.3) and (3.8) are\begingroup%
\renewcommand{\arraystretch}{2.2}%
\begin{equation}
\begin{array}{c}
\displaystyle\nabla ^{D}\cdot \mathbf{E}=0, \\ 
\displaystyle\nabla ^{D}\cdot \mathbf{B}=0, \\ 
\displaystyle\frac{1}{c}\frac{\partial }{\partial t}\mathbf{B}+\mathbf{%
\nabla }^{D}\times \mathbf{E}=\mathbf{0}, \\ 
\displaystyle\frac{1}{c}\frac{\mathbf{\partial E}}{\mathbf{\partial }t}-%
\mathbf{\nabla }^{D}\times \mathbf{B}+\frac{4\pi }{c}\mathbf{J}=\mathbf{0},%
\end{array}%
\end{equation}%
\endgroup Taking derivatives gives the second-order Maxwell's equations for
a fractal\begingroup\renewcommand{\arraystretch}{2.2}%
\begin{equation}
\begin{array}{c}
\displaystyle\mathbf{\nabla }^{D}\cdot (\mathbf{\nabla }^{D}\mathbf{%
E)-\nabla }^{D}(\mathbf{\nabla }^{D}\cdot \mathbf{E})=\frac{1}{c^{2}}\frac{%
\mathbf{\partial }^{2}\mathbf{E}}{\mathbf{\partial }^{2}t}+\frac{4\pi }{c^{2}%
}\frac{\partial }{\partial t}\mathbf{J} \\ 
\text{or} \\ 
\displaystyle\frac{1}{c_{1}^{(p)}}\left[ E_{k},_{p}/c_{1}^{(p)}\right] ,_{p}-%
\frac{1}{c_{1}^{(k)}}\left[ E_{k},_{l}/c_{1}^{(l)}\right] ,_{p}=\frac{1}{%
c^{2}}\frac{\partial ^{2}E_{k}}{\partial t^{2}}+\frac{4\pi }{c}\frac{%
\partial J_{k}}{\partial t}%
\end{array}%
\end{equation}%
\endgroup

Now, consider two special cases:

\textbf{Conductor}. With Ohm's law for anisotropic media (2.22), the
electrodynamics equations simplify to\begingroup\renewcommand{%
\arraystretch}{2.2}%
\begin{equation}
\begin{array}{c}
\nabla ^{D}\cdot \mathbf{D}=0, \\ 
\nabla ^{D}\cdot \mathbf{B}=0, \\ 
\displaystyle\frac{1}{c}\frac{\partial }{\partial t}\mathbf{H}+\mathbf{%
\nabla }^{D}\times \mathbf{E}=\mathbf{0}, \\ 
\displaystyle\frac{\mathbf{\sigma }}{c}\cdot \frac{\mathbf{\partial E}}{%
\mathbf{\partial }t}-\mathbf{\nabla }^{D}\times \mathbf{H}=\mathbf{0},%
\end{array}%
\end{equation}%
\endgroup Upon using the fractal curl\ operation, we obtain a parabolic type
equation\begingroup\renewcommand{\arraystretch}{2.2}%
\begin{equation}
\begin{array}{c}
\displaystyle\mathbf{\nabla }^{D}\cdot \mathbf{\nabla }^{D}\mathbf{H}=\frac{%
\mathbf{\sigma }}{c^{2}}\cdot \frac{\mathbf{\partial H}}{\mathbf{\partial }t}
\\ 
\text{or} \\ 
\displaystyle\frac{1}{c_{1}^{(p)}}\left[ H_{k},_{p}/c_{1}^{(p)}\right] ,_{p}=%
\frac{\sigma _{km}}{c^{2}}\frac{\partial H_{m}}{\partial t}%
\end{array}%
\end{equation}%
\endgroup

\textbf{Dielectric}. From (5.3) we find a hyperbolic type equation\begingroup%
\renewcommand{\arraystretch}{2.2}%
\begin{equation}
\begin{array}{c}
\displaystyle\mathbf{\nabla }^{D}\cdot \mathbf{\nabla }^{D}\mathbf{E}=\frac{%
\mathbf{\sigma }}{c^{2}}\cdot \frac{\mathbf{\partial }^{2}\mathbf{E}}{%
\mathbf{\partial }^{2}t} \\ 
\text{or} \\ 
\displaystyle\frac{1}{c_{1}^{(p)}}\left[ E_{k},_{p}/c_{1}^{(p)}\right] ,_{p}=%
\frac{\sigma _{km}}{c^{2}}\frac{\partial ^{2}E_{m}}{\partial t^{2}}%
\end{array}%
\end{equation}%
\endgroup

\setcounter{equation}{0}

\section{Electromagnetic Energy and Stress}

\textbf{Pontying vector}. Now, we return to (3.3) and (3.7). Multiplying the
first of these equations by $\mathbf{H}$, and the second one by $\mathbf{E}$%
, leads to 
\begin{equation}
\mathbf{\nabla }^{D}\times \mathbf{E}\cdot \mathbf{H}=-\frac{\mathbf{%
\partial B}}{\mathbf{\partial }t}\cdot \mathbf{H}
\end{equation}%
\begin{equation}
\mathbf{\nabla }^{D}\times \mathbf{H\cdot E}=\mathbf{J}\cdot \mathbf{E+}%
\frac{\partial }{\partial t}\mathbf{D}\cdot \mathbf{E}
\end{equation}%
Subtracting (6.2) from (6.1), on account of%
\begin{equation}
\mathbf{H}\cdot \mathbf{\nabla }^{D}\times \mathbf{E}+\mathbf{E}\cdot 
\mathbf{\nabla }^{D}\times \mathbf{H}=\mathbf{\nabla }^{D}\cdot (\mathbf{%
E\times H),}
\end{equation}%
we obtain%
\begin{equation}
\begin{array}{c}
\displaystyle\mathbf{\nabla }^{D}\cdot (\mathbf{E\times H)}+\mathbf{J}\cdot 
\mathbf{E=-}\frac{\partial }{\partial t}\mathbf{D}\cdot \mathbf{E-}\frac{%
\partial }{\partial t}\mathbf{B}\cdot \mathbf{H} \\ 
\text{or} \\ 
\displaystyle e_{ijk}\frac{1}{c_{1}^{(j)}}(E_{k}H_{i}\mathbf{),}%
_{j}+J_{i}E_{i}=-E_{i}\frac{\partial }{\partial t}D_{i}-H_{i}\frac{\partial 
}{\partial t}B_{i}%
\end{array}%
\end{equation}%
Integrating (6.4) over the fractal's volume and applying the Green-Gauss
theorem, yields%
\begin{equation}
\int_{S}\mathbf{G}\cdot \mathbf{n}\text{ }dS_{d}+\int_{\mathcal{W}}\mathbf{J}%
\cdot \mathbf{E}\text{ }dV_{D}=\mathbf{-}\int_{\mathcal{W}}\left[ \frac{%
\partial }{\partial t}\mathbf{D}\cdot \mathbf{E-}\frac{\partial }{\partial t}%
\mathbf{B}\cdot \mathbf{H}\right] dV_{D}
\end{equation}%
where%
\begin{equation}
\mathbf{G}=\mathbf{E}\text{ }\times \mathbf{H}
\end{equation}%
is identified as the Poynting vector, which is seen to have the same form as
in non-fractal media.

Now, the electric and magnetic force densities are\begingroup%
\renewcommand{\arraystretch}{2.2}%
\begin{equation}
\begin{array}{c}
\displaystyle\mathbf{f}^{E}=\rho \mathbf{E}+\mathbf{P}\cdot \mathbf{\nabla }%
^{D}\mathbf{E} \\ 
\displaystyle\mathbf{f}^{M}=\mathbf{J}^{M}\times \mathbf{B}+\mathbf{M}\cdot (%
\mathbf{B}^{D}\mathbf{\nabla })=\frac{1}{c}\mathbf{J}^{E}\times \mathbf{B}+%
\mathbf{M}\cdot (\mathbf{B\nabla }^{D})%
\end{array}%
\end{equation}%
\endgroup This has to be accompanied by an expression of the Amp\`{e}re law
in electrostatic units%
\begin{equation}
\mathbf{\nabla }^{D}\times \mathbf{H}=\frac{4\pi }{c}\mathbf{J}+\frac{1}{c}%
\mathbf{\dot{D}}=\frac{4\pi }{c}(\mathbf{J}+\frac{1}{c}\mathbf{\dot{D}})
\end{equation}

\setcounter{equation}{0}

\section{Conclusion}

We determine the equations governing electromagnetic fields in generally
anisotropic fractal media using the dimensional regularization together with
the recently introduced product measure allowing an independent
characterization of anisotropy in three Cartesian directions of the
Euclidean space in which an anisotropic fractal is embedded. This measure
also gives rise to fractal gradient, divergence and curl operators, which
are shown to satisfy the four fundamental identities of the vector calculus.
Next, the conservation of the electric charge on a fractal shows that the
fractal divergence has to be used. With this Ansatz, in the first place, we
obtain Maxwell equations modified to generally anisotropic fractals using
two independent approaches: a conceptual one (involving generalized Faraday
and Amp\`{e}re laws), and the one directly based on a variational principle
for electromagnetic fields. In both cases the resulting equations are the
same, thereby providing a self-consistent verification of our derivations.

Just as in the previous works of Tarasov, we find that the presence of
anisotropy in the fractal structure leads to a source/disturbance as a
result of generally unequal fractal dimensions in various directions.
However, our modified Mawell equations do not coincide with those of Tarasov
(2006, 2010). The modifications due to fractal geometry carry over to the
parabolic (for conductor) and hyperbolic (for dielectric) equations, while
the Poynting vector is found to have the same form as in the non-fractal
case. Finally, the electromagnetic stress tensor is reformulated for fractal
systems. \ Overall, all the derived relations depend explicitly on three
fractal dimensions $\alpha _{i}$ in the respective Cartesian directions $%
x_{i}$, $i=1,2,3$, as well as the spatial resolution; upon setting all $%
\alpha _{i}=1$, these relations reduce to conventional forms of Maxwell
equations.

\bigskip

\textbf{\Large References}

\bigskip

Falconer, K.:\ \textit{Fractal Geometry: Mathematical Foundations and
Applications}, J. Wiley (2003).

Jumarie, G.:. On the representation of fractional Brownian motion as an
integral with respect to $(dt)^{a}$. \textit{Appl. Math. Lett.} \textbf{18},
739-748 (2005).

Jumarie, G.: Table of some basic fractional calculus formulae derived from a
modified Riemann-Liouville derivative for non-differentiable functions. 
\textit{Appl. Math. Lett.} \textbf{22}(3), 378-385 (2009).

Nottale, L. 2010, Scale relativity and fractal space-time: theory and
applications, \textit{Found. Sci.} \textbf{15}, 101-152.

Li, J. \& Ostoja-Starzewski, M.: Fractal solids, product measures and
fractional wave equations, \textit{Proc. R. Soc. A} \textbf{465}, 2521-2536
(2009) doi:10.1098/rspa.2009.0101; Errata (2010) doi:10.1098/rspa.2010.0491.

Palmer, C. \&\ Stavrinou, P.N. 2004, Equations of motion in a
non-integer-dimensional space, \textit{J. Phys. A:\ Math. Gen.} \textbf{37},
6987-7003.

Seliger, R.L. \&\ Whitham, G.B. 1968, Variational principles in continuum
mechanics. \textit{Proc. R. Soc. A} \textbf{305}, 1-25.

Stillinger, F.H. 1977, Axiomatic basis for spaces with noninteger dimension, 
\textit{J. Math. Phys}. \textbf{18}(6), 1224-1234.

Svozil, K. 1987, Quantum field theory on fractal spacetime: a new
regularisation method, \textit{J. Phys. A:\ Math. Gen.} \textbf{20},
3861-3875.

Tarasov, V.E. 2006 Electromagnetic fields on fractals, \textit{Mod. Phys.
Lett. A} \textbf{21}(20), 1587-1600.

Tarasov, V.E. 2010 \textit{Fractional Dynamics}, Springer\textit{.}

\end{document}